\documentclass[aps, prd, nofootinbib, preprintnumbers, twocolumn]{revtex4-2}
\usepackage{style}

\begin{document}

\title{Absorption of Axion Dark Matter in a Magnetized Medium}

\author{Asher Berlin\,\orcidlink{0000-0002-1156-1482}}
\email{aberlin@fnal.gov}
\affiliation{Theoretical Physics Division, Fermi National Accelerator Laboratory, Batavia, Illinois 60510}

\author{Tanner Trickle\,\orcidlink{0000-0003-1371-4988}}
\email{ttrickle@fnal.gov}
\affiliation{Theoretical Physics Division, Fermi National Accelerator Laboratory, Batavia, Illinois 60510}

\preprint{FERMILAB-PUB-23-168-SQMS-T}
\date{\today}

\begin{abstract}
    Detection of axion dark matter heavier than a meV is hindered by its small wavelength, which limits the useful volume of traditional experiments. This problem can be avoided by directly detecting in-medium excitations, whose $\sim \text{meV} - \text{eV}$ energies are decoupled from the detector size. We show that for any target inside a magnetic field, the absorption rate of electromagnetically-coupled axions into in-medium excitations is determined by the dielectric function. As a result, the plethora of candidate targets previously identified for sub-GeV dark matter searches can be repurposed as broadband axion detectors. We find that a $\text{kg} \cdot \text{yr}$ exposure with noise levels comparable to recent measurements is sufficient to probe parameter space currently unexplored by laboratory tests. Noise reduction by only a few orders of magnitude can enable sensitivity to the QCD axion in the $\sim 10 \ \text{meV} - 10 \ \text{eV}$ mass range.  
\end{abstract}

\maketitle

\noindent
\textbf{Introduction.---} Despite constituting roughly 27\% of the energy density of the universe~\cite{ParticleDataGroup:2020ssz}, the fundamental nature of dark matter (DM) remains elusive. Of the theoretically motivated DM candidates, the QCD axion is particularly remarkable since its existence would also solve the longstanding strong CP problem~\cite{Peccei:1977ur,Peccei:1977hh,Wilczek:1977pj,Weinberg:1975ui}. A generic feature of QCD axion DM models is a coupling between the axion field $a$ and electromagnetism, 
\begin{align}
    \mathcal{L} \supset - \frac{1}{4} \, \gagg \, a \, F^{\mu \nu} \tilde{F}_{\mu \nu} = \gagg \, a \, \E \cdot \B
    ~.
    \label{eq:L_axion}
\end{align}
In the presence of a static magnetic field $\B_0$, the interaction in \Eq{L_axion} converts an axion to an oscillating electromagnetic field~\cite{Sikivie:1983ip}. Directly detecting this field is the underlying principle of many ongoing and planned experiments~\cite{Adams:2022pbo}. Traditional detection schemes utilize cavities with electromagnetic modes resonantly matched to axion masses of $m_a \sim (10^{-6}-10^{-5}) \ \eV$, as motivated by post-inflationary misalignment production and a standard cosmological history~\cite{Abbott:1982af,Turner:1983he,Turner:1985si,Preskill:1982cy,Dine:1982ah}. However, searches across a larger parameter space are motivated by alternative production mechanisms~\cite{Co:2019wyp,Co:2020dya,Co:2019jts,Hagmann:1998me,Gorghetto:2018myk,Battye:1994au,Hindmarsh:2021zkt,Dine:2020pds,Buschmann:2021sdq,Vaquero:2018tib,Buschmann:2019icd,Klaer:2017qhr,Hiramatsu:2010yu,Gorghetto:2020qws,Chang:2019tvx,Harigaya:2019qnl,Co:2017mop,Hindmarsh:2019csc,Daido:2017wwb,Daido:2017tbr,Klaer:2017ond} and axions that couple to the Standard Model similarly to the QCD axion but without the strict connection between coupling strength and mass~\cite{Choi:2009jt,Acharya:2010zx,Chikashige:1980ui,Svrcek:2006yi,Froggatt:1978nt,Co:2020xlh,Witten:1984dg,Arvanitaki:2009fg,Conlon:2006tq,Cicoli:2012sz,Halverson:2017deq}.

Cavities are an exceptional tool to search for axion DM. However, they are fundamentally limited in the axion mass that they can probe. This is because the axion mass must be matched to a resonant frequency of the cavity, which are inversely related to its size. Therefore, to resonantly search for higher axion masses, the cavity must be prohibitively small, limiting the total exposure. Recent strategies to boost exposure to high-mass axions include non-resonant detection of single-photons in a large volume dish antenna~\cite{BREAD:2021tpx}, and modifications to the photon's dispersion relation in dielectric~\cite{Caldwell:2016dcw,Millar:2016cjp,Baryakhtar:2018doz} or plasma~\cite{Lawson:2019brd,Caputo:2020quz,ALPHA:2022rxj} structures tuned to a specific mass.

While these searches are focused on photon detection, another possibility is to directly detect the in-medium excitations in crystal targets involving, e.g.,  electrons~\cite{Hochberg:2016sqx,Krnjaic:2023nxe,Derevianko:2010kz,Catena:2021qsr,Mitridate:2021ctr,Chen:2022pyd,Catena:2022fnk,Blanco:2019lrf,Knapen:2021run,Hochberg:2015fth,Hochberg:2016ajh,Hochberg:2017wce,Hochberg:2021pkt,Geilhufe:2019ndy,Essig:2011nj,Derenzo:2016fse,Essig:2015cda,Coskuner:2019odd,Hochberg:2016ntt,Catena:2023qkj,Catena:2023awl,Du:2022dxf,Griffin:2021znd,Trickle:2022fwt,Griffin:2019mvc,Trickle:2019nya,Graham:2012su,Inzani:2020szg,Knapen:2021bwg}, phonons~\cite{Cox:2019cod,Mitridate:2020kly,Knapen:2017ekk,Coskuner:2021qxo,Griffin:2018bjn,Trickle:2020oki,Griffin:2019mvc,Trickle:2019nya,Marsh:2022fmo}, and magnons~\cite{Barbieri:1985cp,Flower:2018qgb,Mitridate:2020kly,Chigusa:2020gfs,Trickle:2019ovy,Esposito:2022bnu}. Since the energy of these modes ($\sim \meV - \eV$) is not set by the target size, but rather by the physics of the local environment, they are ideal for high-mass axion searches. Furthermore, the manufacturing of low-noise targets and the technology required to detect single quanta of such excitations is at the forefront of the DM direct detection community and is thus an active area of development~\cite{Essig:2022dfa}. In particular, current experiments, such as CDEX~\cite{CDEX:2022kcd}, DAMIC~\cite{DAMIC:2019dcn,DAMICM:2023gxo,DAMIC:2016qck,DAMIC:2020cut,Settimo:2020cbq,DAMIC:2015znm}, EDELWEISS~\cite{EDELWEISS:2020fxc,EDELWEISS:2018tde,EDELWEISS:2019vjv}, SENSEI~\cite{SENSEI:2019ibb,SENSEI:2020dpa,Crisler:2018gci}, and SuperCDMS~\cite{SuperCDMS:2020ymb,SuperCDMS:2018mne,CDMS:2009fba}, utilize eV-scale electronic excitations in Si and Ge targets. More novel targets with sub-eV electronic excitations have also been proposed, such as narrow gap semiconductors~\cite{SuperCDMS:2022kse}, Dirac materials~\cite{Hochberg:2015fth,Hochberg:2017wce,Geilhufe:2019ndy,Coskuner:2019odd,Hochberg:2016ntt,Catena:2023qkj,Catena:2023awl}, spin-orbit coupled materials~\cite{Chen:2022pyd,Inzani:2020szg}, and doped semiconductors~\cite{Du:2022dxf}. Additionally, phonon excitations have been studied in a wide variety of target materials~\cite{Mitridate:2020kly,Knapen:2017ekk,Coskuner:2021qxo,Griffin:2018bjn,Griffin:2019mvc,Knapen:2021bwg,Griffin:2020lgd}, including GaAs and \ce{Al2O3} as planned for the TESSARACT experiment~\cite{Chang2020}.

\begin{figure*}
    \centering
    \includegraphics[width=0.75\textwidth]{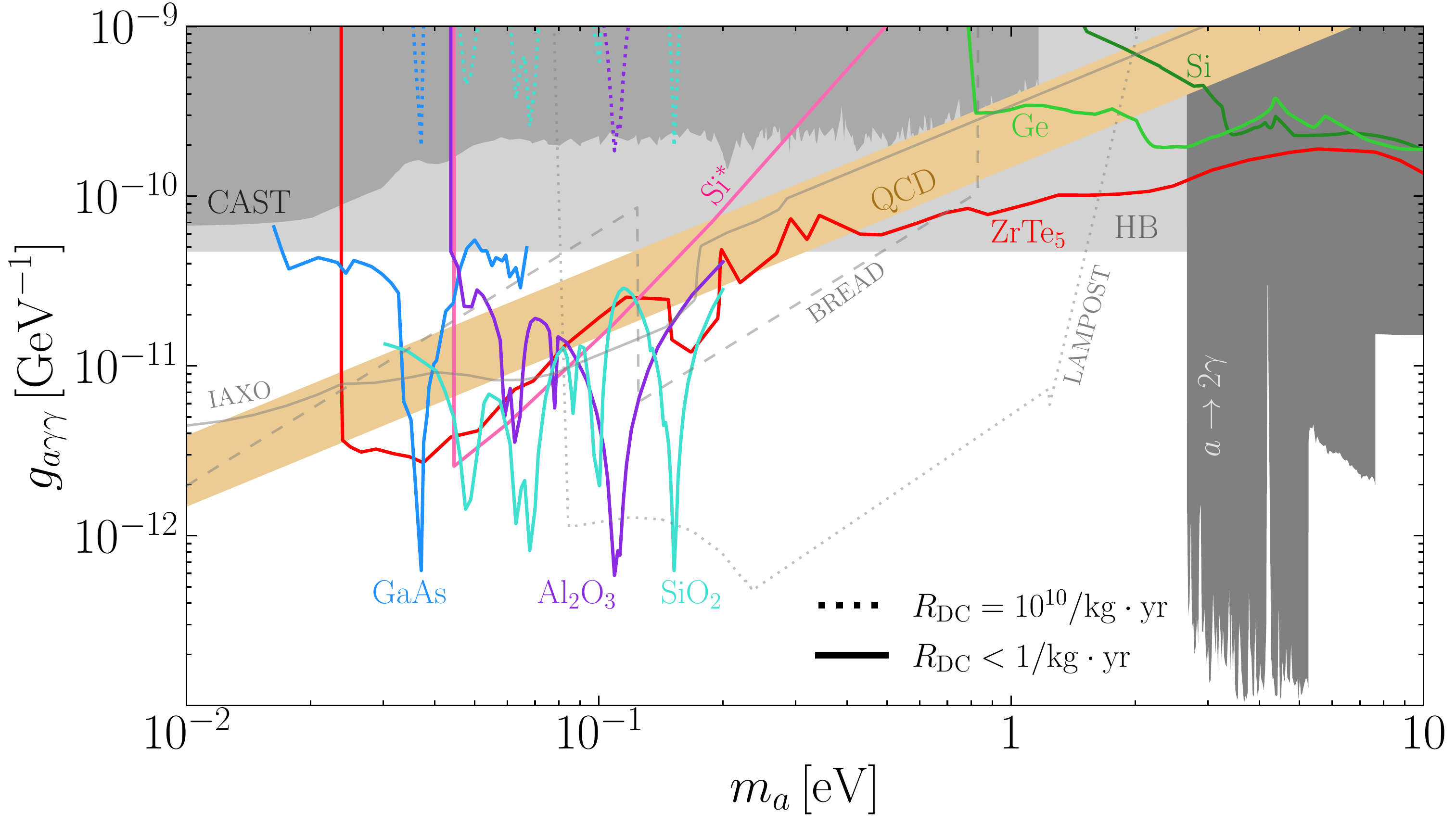}
    \caption{Projected sensitivity to electromagnetically-coupled axion dark matter, for a $\kgyr$ exposure of various targets inside a $10 \ \text{T}$ magnetic field. Solid colored lines assume negligible backgrounds, and dotted colored lines assume a dark count (DC) rate of $R_\text{DC} = 10^{10}/\kgyr$. Targets utilizing single-phonon excitations include GaAs (light blue) and  \ce{Al2O3} (purple) (proposed for the TESSARACT experiment~\cite{Chang2020}), as well as \ce{SiO2} (turquoise). 
    The Si (dark green) and Ge (light green) targets correspond to single-electron excitations currently searched for by the CDEX~\cite{CDEX:2022kcd}, DAMIC~\cite{DAMIC:2019dcn,DAMICM:2023gxo,DAMIC:2016qck,DAMIC:2020cut,Settimo:2020cbq,DAMIC:2015znm}, EDELWEISS~\cite{EDELWEISS:2020fxc,EDELWEISS:2018tde,EDELWEISS:2019vjv}, SENSEI~\cite{SENSEI:2019ibb,SENSEI:2020dpa,Crisler:2018gci}, and SuperCDMS~\cite{SuperCDMS:2020ymb,SuperCDMS:2018mne,CDMS:2009fba} experiments. Doped Si (Si$^*$, pink)~\cite{Du:2022dxf} and \ce{ZrTe5} (red)~\cite{Chen:2022pyd} correspond to novel targets utilizing low-energy electronic excitations. Shaded gray regions are existing limits derived from horizontal branch (HB) star cooling~\cite{Dolan:2022kul}, the CAST helioscope~\cite{CAST:2007jps,CAST:2017uph}, and searches for $a \to 2 \gamma$ decays by the MUSE~\cite{Regis:2020fhw}, HST~\cite{Carenza:2023qxh}, and VIMOS~\cite{Grin:2006aw} telescopes. Also shown as gray lines are projections from the IAXO (solid)~\cite{Shilon:2012te}, BREAD (dashed) (assuming DCs of $\sim 10^4$ or $\lesssim 1$ for masses below or above $\sim 100 \ \meV$, respectively, over $10^3 \ \text{days}$)~\cite{BREAD:2021tpx}, and LAMPOST (dotted) (assuming $\sim 10$ DCs per $10^6 \ \text{s}$ run and $\sim 10^4$ runs)~\cite{Baryakhtar:2018doz} experiments. The orange band denotes the range of couplings and masses as motivated by the QCD axion.}
    \label{fig:projected_reach_optimistic}
\end{figure*}

In this \textit{Letter}, we demonstrate that the entirety of these ideas can be used to search for the axion-photon coupling in \Eq{L_axion}, provided that the target can be placed inside a magnetic field, thus creating a ``magnetized medium." In particular, we show that in a magnetized medium the inclusive axion absorption rate into in-medium excitations is directly related to the dielectric function. While certain signals of axion DM have previously been found to be related to the dielectric function on a case-by-case basis, we show here that this is universal. This is important both experimentally, since the dielectric can be measured, and theoretically, because it broadly captures the absorption rate into any in-medium excitation, abstracting away from calculations specific to any single  excitation. This allows us to easily evaluate the sensitivity of various materials, as well as identify a larger scope of relevant signals that have previously been overlooked, such as low-energy electronic excitations. Below, we begin by deriving the absorption rate with two methods. The first derivation involves self-energies, analogous to calculations performed in the context of direct detection experiments; the second is provided within the language of classical axion electrodynamics. These derivations provide complementary ways to understand the underlying physics. We then discuss the projections shown in \Fig{projected_reach_optimistic}, which illustrate the promising ability to explore new, high-mass, QCD axion parameter space.

\vspace{0.5cm}
\noindent
\textbf{Absorption Rate.---}
Before deriving the rate for axion absorption in a magnetized medium, we begin with a synopsis of the final result for isotropic targets. The total axion absorption rate, per unit exposure, is given by
\begin{align}
    \Rsig \simeq \bigg( \frac{\gagg \, B_0}{m_a} \bigg)^2 \, \frac{\rhodm}{\rhoT} ~ \text{Im}\bigg[ \frac{-1}{\eps (m_a)} \bigg]
    ~,
    \label{eq:main_rate}
\end{align}
where $\rhodm \simeq 0.4 \ \GeV/\cm^3$ is the local axion DM energy density, $\rhoT$ is the mass density of the target, and $\eps(m_a)$ is the dielectric function evaluated at energy $\w = m_a$ and momentum deposition $q = 0$, appropriate for absorption kinematics ($q \ll \omega$) which are assumed throughout. The simplicity of this expression derives from the separability of the axion absorption process as $a + B_0 \to E$ followed by absorption of the corresponding electric field.\footnotemark \addtocounter{footnote}{-1} The former process is governed by the strength of the external magnetic field and $\gagg$, while the latter is determined by the dielectric function, independent of both the axion physics and, in the $q \ll \w$ limit, the magnetic permeability. This separability is advantageous since, in principle, the dielectric function of the target can be measured directly. In the absence of measurement, this parameterization is useful as a bridge between particle physics and first principles condensed matter calculations. First principles calculations are a useful tool to understand the contributions of individual excitations, which cannot be understood from a measurement of the inclusive dielectric function. However, this generally ceases to be a problem when the various excitations are sufficiently separated in energy.

The idea of relating the dielectric function to the DM absorption rate into in-medium excitations has been used for other DM models~\cite{Hochberg:2016sqx,Krnjaic:2023nxe,Derevianko:2010kz,Mitridate:2021ctr,Inzani:2020szg,Knapen:2021bwg},
as well as in calculations of the DM absorption rate into in-medium photon states~\cite{Gelmini:2020kcu,Caputo:2020quz,ALPHA:2022rxj}.
For example, for kinetically-mixed dark photon DM, $\Ap$, the absorption rate 
into in-medium excitations 
is~\cite{Hochberg:2016sqx,Mitridate:2021ctr,Knapen:2021bwg} 
\begin{align}
    \Rsig \simeq \kappa^2 \, \frac{\rhodm}{\rhoT} ~ \text{Im}\bigg[ \frac{-1}{\eps (\mAp)} \bigg]
    ~,
    \label{eq:rate_dp}
\end{align}
where $\kappa$ is the kinetic-mixing parameter and $\mAp$ is the $\Ap$ mass. The similarity between the dark photon absorption rate in \Eq{rate_dp} and the axion absorption rate in \Eq{main_rate} is immediately clear. As a result, for DM particles of the same mass, the sensitivity to electromagnetically-coupled axions can be simply rescaled via the mapping $\gagg \, B_0 \leftrightarrow \kappa \, m_a$, corresponding to
\begin{align}
    \gagg \sim 10^{-10} \ \GeV^{-1} \times \bigg( \frac{\kappa}{10^{-14}} \bigg) \,  \, \bigg( \frac{m_a}{\meV} \bigg) \bigg( \frac{\text{T}}{B_0} \bigg)
    ~.
    \label{eq:constraint_rescale}
\end{align}

We now turn to the details of the calculation, beginning with a self-energy treatment similar to Refs.~\cite{Hardy:2016kme,Mitridate:2021ctr,Krnjaic:2023nxe}, then turning to an alternative derivation employing classical equations of motion, similar to Refs.~\cite{Dubovsky:2015cca,Marsh:2022fmo}.

\footnotetext{This description is appropriate for the axion masses of interest here, such that the target size is much larger than the skin-depth $\sim \big(m_a \, \text{Im} \left[\sqrt{\eps}\right] \big)^{-1}$.
In this case, the effects of boundary conditions (arising from, e.g., the finite size of the target or external electromagnetic shielding) can be ignored~\cite{Chaudhuri:2014dla,ALPHA:2022rxj,Balafendiev:2022wua}.
} 

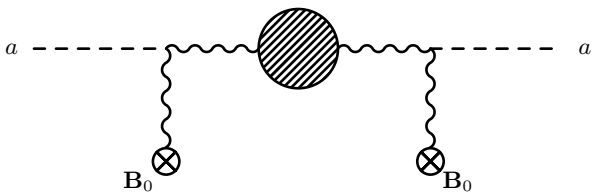
\begin{figure} 
    \vspace{-2.5em}
    \centering
    \begin{fmffile}{pi_aa}
        \begin{fmfgraph*}(200,80)
            %%%
            % https://tex.stackexchange.com/questions/175659/circle-with-a-cross-inside-feynmp   
            \fmfcmd{
                path quadrant, q[], otimes;
                quadrant = (0, 0) -- (0.5, 0) & quartercircle & (0, 0.5) -- (0, 0);
                for i=1 upto 4: q[i] = quadrant rotated (45 + 90*i); endfor
                otimes = q[1] & q[2] & q[3] & q[4] -- cycle;
            }
            \fmfwizard
            %%%
            \fmfleft{i} \fmfright{f}
            \fmfbottom{b1,b2}
            \fmf{dashes}{i,m1}
            \fmf{photon}{m1,v}
            \fmf{photon}{v,m2}
            \fmfblob{.15w}{v}
            \fmf{dashes}{m2,f}
            \fmffreeze
            \fmfforce{vloc(__m1) shifted (0,-15mm)}{b1}
            \fmfforce{vloc(__m2) shifted (0,-15mm)}{b2}
            \fmf{photon}{b1,m1}
            \fmf{photon}{b2,m2}
            \fmfv{d.sh=otimes,d.f=empty,d.si=.05w}{b1}
            \fmfv{d.sh=otimes,d.f=empty,d.si=.05w}{b2}
            \fmflabel{$a$}{i}
            \fmflabel{$a$}{f}
            \fmflabel{$\B_0$}{b1}
            \fmflabel{$\B_0$}{b2}
        \end{fmfgraph*}
    \end{fmffile}
    \vspace{1em}
    \caption{An illustrative Feynman diagram for the axion $a$ self-energy in a magnetized medium. The optical theorem relates the imaginary part of this diagram to the axion absorption rate. The external magnetic field $\B_0$ is represented as a background source, and the shaded ``blob" represents \textit{any} in-medium modifications to the photon propagator (e.g., from electron and phonon excitations).}
    \label{fig:feynman_diagram}
\end{figure}

\vspace{0.5cm}
\noindent
\textit{Self-Energy Calculation.---} 
The starting point of the self-energy derivation is the optical theorem, which relates the absorption rate of a particle to the imaginary part of its self-energy. The self-energy can then be computed by summing over all the relevant Feynman diagrams. The Feynman diagram for axion absorption in a magnetized medium is shown in Fig.~\ref{fig:feynman_diagram}; the vertex Feynman rule is derived from the Lagrangian in \Eq{L_axion}, and the photon propagator is modified due to the presence of the medium.  

Following Refs.~\cite{Hardy:2016kme,Mitridate:2021ctr,Krnjaic:2023nxe}, instead of directly computing the diagram in Fig.~\ref{fig:feynman_diagram}, we identify the diagrams which mix the axion and photon, and those which do not. If the axion and photon mix, the in-medium fields are those that diagonalize the $2\times2$ self-energy matrix of $a$ and $A$. The diagonalizing fields $\hat{a}$ and $\hat{A}$ (the axion-like and photon-like fields, respectively) are a linear combination of $a$ and $A$. The probability of axion absorption per unit time is then related to the self-energy of $\hat{a}$, 
\begin{align}
    \Gamma \simeq - \frac{1}{m_a} \, \text{Im} \bigg[ \, \Pi_{aa} + \sum_{\lambda} \frac{\Pi_{aA}^\lambda \, \Pi_{Aa}^\lambda}{m_a^2 - \Pi_{AA}^\lambda} \, \bigg]
    ~, 
    \label{eq:abs_gamma_QM}
\end{align}
written in terms of the unmixed self-energies of the $a, A$ fields, $\Pi_{aa}$ and $\Pi_{AA}$, respectively, and the mixing terms, $\Pi_{aA}$, $\Pi_{Aa}$. $\Pi_{AA}^\lambda$ is the self-energy of $A$ projected onto the $\lambda^\text{th}$ polarization $e^\lambda_\mu$, defined to diagonalize $\Pi_{AA}^{\mu\nu}$: $\Pi_{AA}^\lambda = - e^\lambda_\mu \, \Pi_{AA}^{\mu \nu} \, e^\lambda_\nu$. $\Pi_{aA}^\lambda$ is the self-energy mixing $a$ and $A$, projected onto the same photon polarization vector, i.e., $\Pi_{aA}^\lambda = -e^\lambda_\mu \, \Pi_{aA}^\mu$. 

In the absence of direct axion couplings to electrons, the only non-zero self-energies (ignoring vacuum processes) are $\Pi_{AA}^{\mu \nu}$ and $\Pi_{aA}^\mu$. Furthermore, since the Ward identities ($Q_\mu \Pi_{AA}^{\mu \nu} = Q_\mu \Pi_{aA}^\mu = 0$, where $Q^\mu = (\w, \mathbf{q})$) relate the temporal and spatial components, only $\Pi_{AA}^{ij}$ and $\Pi_{aA}^i$ need to be computed. The photon self-energy is determined by the dielectric tensor $\eps^{ij}$ through $\Pi_{AA}^{ij} = -\w^2 (1 - \eps^{ij})$. Note that the spatial component of the photon polarization vectors diagonalize the dielectric tensor, $\eps^{ij} = \sum_\lambda \eps_\lambda e^i_\lambda e^j_\lambda$, where $\eps_\lambda \equiv e^\lambda_i \, \eps^{ij} \, e^\lambda_j$, such that $\Pi^\lambda_{AA} \simeq \w^2 (1 - \eps_\lambda)$. The axion-photon self-energies are determined by the term in \Eq{L_axion} involving the vector potential which, after integrating by parts, is $\gagg \, \dot{a} \, \A \cdot \mathbf{B}_0$. The mixed self-energies are then given by $\Pi_{aA}^i = i \gagg \, m_a \, B_0^i = - \Pi_{Aa}^i$. Substituting $\Pi_{AA}$ and $\Pi_{a A}$ into \Eq{abs_gamma_QM}, and taking the $q \to 0$ limit, results in
\begin{align}
    \Gamma \simeq \frac{\gagg^2}{m_a} ~ \sum_\lambda \left( \mathbf{e}_\lambda \cdot \B_0 \right)^2 \, \text{Im} \bigg[ \frac{-1}{\eps_\lambda(m_a)} \bigg] ~.
    \label{eq:Gamma_QM}
\end{align}

This expression can be further simplified in the limit of an isotropic target. The dielectric of an isotropic target is independent of polarization, $\eps_\lambda = \eps$, and the photon polarization vectors are the standard transverse $e^\mu_\pm  = (0, \hat{\mathbf{q}}_\pm)$ and longitudinal $e^\mu_L = (q, \w \hat{\mathbf{q}})/\sqrt{Q^2}$ ones, where $\hat{\mathbf{q}}_\pm$ are two vectors mutually orthonormal to $\hat{\mathbf{q}}$. The sum over polarizations can be performed using $\sum_{\lambda} e^i_\lambda e^j_\lambda = \delta^{ij}$. Applying these approximations, \Eq{Gamma_QM} simplifies to
\begin{align}
    \Gamma \simeq \frac{(\gagg \, B_0)^2}{m_a} ~ \text{Im} \bigg[ \frac{-1}{\eps(m_a)} \bigg]
    ~.
\end{align}
The rate per unit exposure $R$ in \Eq{main_rate} is then obtained by multiplying $\Gamma$ by the number of axions in the target and dividing by the target mass.

\vspace{0.5cm}
\noindent
\textit{Classical Equations of Motion.---}
Alternatively, the absorption rate can be derived using classical axion electrodynamics. Throughout, we will implicitly work in Lorenz gauge. Our starting point is the wave equation for the vector potential, which is approximately $\epsv \, \partial_t^2 \A \simeq \jv_a$, where $\jv_a = \gagg \, \dot{a} \, \B_0$ is the axion effective current and $\epsv$ is the dielectric tensor. This equation is trivially solved for $\A$ by switching to momentum space and projecting onto the photon polarization vectors $\mathbf{e}_\lambda$, which determines the corresponding electric field to be
\begin{align}
    \E \simeq - \gagg \,  a \,  \sum_{\lambda} \, \frac{( \mathbf{e}_\lambda \cdot \mathbf{B}_0 )}{\eps_\lambda} ~ \mathbf{e}_\lambda
    ~.
    \label{eq:Eaxion}
\end{align}
The rate for axion absorption is governed by the axion equation of motion, which is approximately
\begin{align}
\label{eq:axion4}
(\partial_t^2 + m_a^2) \, a \simeq \gagg \, \E \cdot \B_0
~,
\end{align}
with $\E$ given by \Eq{Eaxion}. The probability per unit time for axion absorption is determined by solving for the imaginary component of the axion frequency in \Eq{axion4}, $\Gamma \simeq \text{Im}(- \omega^2) / m_a$. This leads to a result in agreement with \Eq{Gamma_QM}.\footnotemark \addtocounter{footnote}{-1}

\vspace{0.5cm}
\noindent
\textbf{Projected Sensitivity.---} 
\footnotetext{\label{foot:dp}A similar calculation yields the kinetically-mixed dark photon absorption rate. To leading order in $\kappa$, the approximate equation of motion for the visible vector potential is $\epsv \, \partial_t^2 \A \simeq - \kappa \, \mAp^2 \A^\p$, where we have employed the visible/invisible field basis in which $\A$ and $\A^\p$ are coupled and decoupled from Standard Model sources, respectively (see, e.g., Refs. \cite{Graham:2014sha,Berlin:2023mti}). Analogous to the axion case, the $\A$ equation of motion can be solved for in momentum space, and then substituted into the equation of motion for $\A^\p$, $(\partial_t^2 + \mAp^2) \, \A^\p \simeq - \kappa \, \mAp^2 \A$, whose imaginary frequency component determines the absorption rate of \Eq{rate_dp}.}
The sensitivity of a variety of targets is shown in \Fig{projected_reach_optimistic}, assuming a $\kgyr$ exposure and $B_0 = 10 \ \text{T}$. In our estimates, we demand $N > 3 \, \sqrt{1 + N_\text{DC} + \delta^2 N_\text{DC}^2}\, $, where $N = R \, M_\text{T} \, t$ and $N_\text{DC} = R_\text{DC} \, M_\text{T} \, t$ are the the number of signal events and dark counts (DCs), respectively, $R_\text{DC}$ is the DC rate, $M_\text{T}$ is the target mass, $t$ is the exposure time, and $\delta \leq 1$ is the systematic uncertainty in the DC. In the absence of background, $N_\text{DC} \ll 1$, the sensitivity to the axion-photon coupling scales as $\gagg \propto (M_\text{T} \, t)^{-1/2}$. If backgrounds are instead significant, $N_\text{DC} \gg 1$, the reliance of the signal on $B_0$ allows these DCs to be directly measured by removing the magnetic field, thereby suppressing systematic uncertainties, $\delta \ll 1$. In the statistically-limited regime, $\delta \ll 1 / \sqrt{N_\text{DC}}$, an observable signal only needs to overcome Poisson fluctuations in noise, such that $\gagg \propto (R_\text{DC} / M_\text{T} \, t)^{1/4}$. Below, we begin by discussing the sensitivity of various targets assuming negligible backgrounds, and then proceed to examine the impact of currently measured noise sources.

For $m_a \gtrsim \eV$, enough energy is deposited to excite an electron across the $\sim 1 \ \eV$ band gap in standard semiconductors, such as Si and Ge, which is then read out by drifting the charge to a sensing output. This is the operating principle of many ongoing experiments, such as CDEX~\cite{CDEX:2022kcd}, DAMIC~\cite{DAMIC:2019dcn,DAMICM:2023gxo,DAMIC:2016qck,DAMIC:2020cut,Settimo:2020cbq,DAMIC:2015znm}, EDELWEISS~\cite{EDELWEISS:2020fxc,EDELWEISS:2018tde,EDELWEISS:2019vjv}, SENSEI~\cite{SENSEI:2019ibb,SENSEI:2020dpa,Crisler:2018gci}, and SuperCDMS~\cite{SuperCDMS:2020ymb,SuperCDMS:2018mne,CDMS:2009fba}. In our estimate of the signal rate in \Eq{main_rate}, we use the measured dielectric functions of Si and Ge from Ref.~\cite{1985}, and do not incorporate multi-phonon responses at lower energies~\cite{Knapen:2021bwg} since these are subdominant to the single-phonon responses of the polar materials discussed below. As shown in \Fig{projected_reach_optimistic}, background-free Ge targets have the potential to be the best laboratory-based search for the QCD axion for masses greater than that probed by the CAST helioscope~\cite{CAST:2007jps,CAST:2017uph} and smaller than that probed by astrophysical searches for $a \to 2 \gamma$ decays~\cite{Regis:2020fhw,Carenza:2023qxh,Grin:2006aw}. 

While the energy of electronic excitations is limited to $\sim \eV$ scales in standard semiconductors, novel targets have lower $\sim \meV$ electronic excitations. For example, Dirac~\cite{Hochberg:2015fth,Hochberg:2017wce,Geilhufe:2019ndy,Coskuner:2019odd,Hochberg:2016ntt,Catena:2023qkj,Catena:2023awl} and spin-orbit coupled materials~\cite{Chen:2022pyd,Inzani:2020szg} have small bulk band gaps. One such target that falls under both categories is \ce{ZrTe5}; while its Dirac character is somewhat debated~\cite{Chen:2022pyd}, the presence of its small band gap has been firmly established~\cite{PhysRevB.94.081101}. However, since its dielectric response has not been accurately measured, we adopt the first-principles calculation performed in Ref.~\cite{Chen:2022pyd}. In addition to pure targets, doping is another method to create electronic states below the band gap. Recently, Ref.~\cite{Du:2022dxf} studied Si doped with phosphorus as a candidate target, using an analytic model for the dielectric response consistent with  measurements~\cite{PhysRevB.52.16486, PhysRevLett.71.3681}. In \Fig{projected_reach_optimistic}, we rescale their quoted dark photon sensitivity, using the mapping described above. 

Phonon excitations~\cite{Mitridate:2020kly,Knapen:2017ekk,Coskuner:2021qxo,Griffin:2018bjn,Griffin:2019mvc,Knapen:2021bwg,Griffin:2020lgd} in the $\sim (1 -100) \ \meV$ energy range have also been studied as an avenue to detect axions~\cite{Mitridate:2020kly,Marsh:2022fmo}. The results shown in \Fig{projected_reach_optimistic} are consistent with the first principles calculation done in Ref.~\cite{Mitridate:2020kly}. We have chosen to focus on \ce{GaAs}, \ce{Al2O3}, and \ce{SiO2} targets, as they are of active investigation in the sub-GeV DM community; the first two are currently planned for the TESSARACT experiment~\cite{Chang2020}, and \ce{SiO2} has also been identified as an optimal target for light DM scattering~\cite{Griffin:2019mvc}. The measured dielectric data for these targets in the phonon energy regime is taken from Ref.~\cite{Knapen:2021bwg}. 

\vspace{0.5cm}
\noindent
\textit{Backgrounds.---}
The sensitivity of detectors focused on readout of single-electron excitations are currently limited by DCs~\cite{Essig:2022dfa}. The SENSEI experiment~\cite{SENSEI:2021hcn,SENSEI:2020dpa}, utilizing a Si Skipper CCD detector, is the current state-of-the-art for measuring charge from single-electron excitations. Recent measurements indicate a DC rate of $10^8 / \kgyr$ and $10^6 / \kgyr$ for energy ranges of $\lesssim 4.7  \ \eV$ and $\sim (4.7 - 8.3) \ \eV$, respectively, and rates consistent with zero for larger energies. Assuming that similar background levels in the two-electron bin can be achieved in a Ge based detector, noise reduction by three orders of magnitude  ($R_\text{DC} \sim 10^3/\kgyr)$ is necessary to attain sensitivity to the QCD axion at eV-scale masses. Similar noise levels, but at much lower energies, are also necessary for doped Si and \ce{ZrTe5} targets to attain sensitivity to the QCD axion at $\sim 100 \ \meV$ masses.

Calorimetric detectors (e.g., SuperCDMS CPD~\cite{SuperCDMS:2020aus}, which measures phonons produced from single-electron excitations) have registered significantly higher noise levels, $R_\text{DC} \sim 10^{10} / \kgyr$. As an estimate of the backgrounds that will contaminate sensors based on detection of single-phonon excitations, we assume $R_\text{DC} = 10^{10}/\kgyr$ and $\delta \lesssim 1 / \sqrt{N_\text{DC}}$ for the dotted colored lines shown in \Fig{projected_reach_optimistic}. If the DCs are reduced to just $\sim 10^8 / \kgyr$ at $\sim 100 \ \meV$ energies, an \ce{Al2O3} or \ce{SiO2} target can be sensitive to the QCD axion at couplings smaller than that bounded by considerations of stellar energy loss. 

While their origin is unknown, noise levels are generally peaked towards smaller energies~\cite{Fuss:2022fxe}. Recent work has focused on progressing the understanding of such backgrounds. For instance, it has been calculated that a subdominant fraction of DCs arises from secondary radiation generated by high-energy tracks~\cite{Du:2020ldo} and photons~\cite{Berghaus:2021wrp} in low-threshold charge and phonon detectors, respectively. Additionally, a bulk of eV-scale phonon backgrounds in superconducting calorimetric detectors has recently been shown to emerge from cooling-induced micro-fractures in auxiliary detector components~\cite{Anthony-Petersen:2022ujw}. 

The setup investigated in our work differs from previously proposed applications of these targets in the use of a large external magnetic field, whose dominant effect will be disrupting detection technology based on superconducting devices (measurements of typical semiconducting sensors have found minor changes to their operation when exposed to $\sim 1 \ \text{T}$ magnetic fields~\cite{CCDBfield}). For example, transition edge sensors (TES) (the main technology for reading out single-phonon excitations in the TESSARACT experiment) are not operative in $\gtrsim \mu \text{T}$ magnetic fields~\cite{deWit:2022ckf}. However, there is a strong dependence on the direction of the magnetic field relative to the face of the TES. Additionally, if the region of bulk target within the external magnetic field can be physically separated from the superconducting detector, these problems can be avoided. Other complications arising from, e.g., additional radioactive components or Lorentz-force induced mechanical stress may also be introduced. A detailed investigation of these effects is beyond the scope of this work and these will also need to be confronted in other axion experiments operating in the $\meV - \eV$ energy range~\cite{Baryakhtar:2018doz,BREAD:2021tpx}. However, we do not expect fundamental roadblocks in this approach since $\sim 10 \ \text{T}$ magnetic fields only change electronic energies at the level of $\sim \meV$, well below the energies investigated here. 

\vspace{0.5cm}
\noindent
\textbf{Discussion.---}
The past decade has seen a meteoric rise in target proposals to hunt for sub-GeV DM. In an external magnetic field these targets are also powerful probes of axions in a mass range that is notoriously difficult to explore, corresponding to $m_a \sim 10 \ \meV - 10 \ \eV$. Searching for axions with these targets has the intrinsic advantage of directly utilizing all future experimental improvements in background reduction, an effort which has assiduously driven direct detection experiments to incredible precision in recent history. The axion absorption rate in a magnetized medium can be simply written in terms of the measurable dielectric function, encoding all in-medium responses. This synergizes with future developments towards optimizing the energy loss function of the material, $\text{Im}(-1/\eps)$, as well as further study into other novel low-energy excitations, such as axion quasiparticles~\cite{Marsh:2018dlj,Schutte-Engel:2021bqm} or chiral phonons~\cite{Romao:2023zqf}. 

\vspace{0.5cm}
\noindent 
\textbf{Acknowledgements.}
\textit{
    We would like to thank Aaron Chou, Juan Estrada, Roni Harnik, Yoni Kahn, Alex Millar, Tongyan Lin, Andrea Mitridate, Kris Pardo, and Kathryn Zurek for helpful conversations, and Ciaran O'Hare for the compilation of axion limits in Ref.~\cite{AxionLimits}. This material is based upon work supported by the U.S. Department of Energy, Office of Science, National Quantum Information Science Research Centers, Superconducting Quantum Materials and Systems Center (SQMS) under contract number DE-AC02-07CH11359. Fermilab is operated by the Fermi Research Alliance, LLC under Contract DE-AC02-07CH11359 with the U.S. Department of Energy.
}

\bibliographystyle{utphys3}
\bibliography{biblio}

\end{document}